\documentstyle[multicol,prl,aps,epsf]{revtex}
\begin{document}

\title{Phase ordering and roughening on growing films}

\author{Barbara Drossel$^{1,2}$ and Mehran Kardar$^3$}
\address{${}^1$ Department of Physics, 
  University of Manchester, Manchester M13 9PL, England}
\address{${}^2$ School of Physics and Astronomy, Raymond and Beverley Sackler Faculty of Exact Sciences, Tel Aviv 69978, Israel}
\address{${}^3$ Department of Physics, Massachussetts Institute of Technology, 
Cambridge, MA02139, USA}
 
\maketitle

\begin{abstract}
We study the interplay between surface roughening 
and phase separation during the growth of binary films. 
Already in 1+1 dimensions, we find a variety of different scaling behaviors, 
depending on how the two phenomena are coupled. 
In the most interesting case, related to the advection of a 
passive scalar in a velocity field, 
nontrivial scaling exponents are obtained in simulations. 

\end{abstract}

\begin{multicols}{2}

Thin solid films are grown for a variety of technological applications, 
using molecular beam epitaxy (MBE) or vapor deposition. 
In order to create materials with specific electronic, optical, 
or mechanical properties, often more than one type of particle is deposited. 
When the particle mobility in the bulk is small, surface configurations 
become frozen in the bulk, leading to anisotropic structures that reflect the
growth history, and are different from bulk equilibrium phases\cite{hel95}. 
Characterizing structures generated during composite film growth is not
only of technological importance, but represents also an interesting
and challenging problem in statistical physics. 

In this paper, we examine the growth of binary films through vapor deposition, 
and study some of the rich phenomena resulting from the interplay of {\it phase 
separation} and {\it surface roughening}. 
Simple models for {\it layer by layer} growth assume either that the 
probability that an incoming atom sticks to a given surface site depends 
on the state of the neighboring sites in the layer below \cite{kan90}, 
or that the top layer is fully thermally equilibrated \cite{dro97}. 
Assuming that the bulk mobility is zero, once a  site is occupied, 
its state does not change any more. 
If the growth rules are invariant under the exchange of the two particle types, 
the phase separation is in the universality class of an 
equilibrium Ising model. 
Correlations perpendicular to the growth direction are
characterized by the critical exponent $\nu$ of the Ising model, and
those parallel to the growth direction by the exponent $\nu z_m$, with $z_m$
being the dynamical critical exponent of the Ising model.

However, the {\it layer by layer} growth mode underlying these simple
models is unstable, and the growing surface becomes {\it rough}. 
In many cases the fluctuations in the height $h({\bf x},t)$, at position
${\bf x}$ and time $t$ are {\it self-affine}, with correlations
\begin{equation}\label{Chh}
\langle \left[h({\bf x},t)-h({\bf x'},t')\right]^2\rangle 
\sim |{\bf x-x'}|^{2\chi} g\left(t/|{\bf x-x'}|^{z_h}\right),
\end{equation}
where $\chi$ is the roughness exponent,  
and $z_h$ is a dynamical scaling exponent.
A computer model with local sticking probabilities that allows for a rough 
surface was introduced in  \cite{kot98}.
In 1+1 dimensions, the authors find phase separation into domains
(with sizes consistent with the Ising model), and a very rough surface
profile with sharp minima at the domain boundaries.
We may ask the following questions:
{\bf (1)} Are the roughness exponents different at the phase transition point?
{\bf (2)} Are the critical exponents modified on a rough surface?
We shall demonstrate that the coupling of roughening and phase 
separation leads to a rich phase diagram, and to nontrivial critical exponents 
already in 1+1 dimensions.

To characterize phase separation, we introduce an {\it order parameter}
$m({\bf x},t)$, which is the difference in the densities of the two
particle types at the surface at position ${\bf x}$ and time $t$. 
The interplay between the fluctuations in $m$, and the height $h$ is
captured  phenomenologically by the coupled Langevin equations, 
\begin{eqnarray}
\partial_t h &=& \nu \nabla^2 h + {\lambda\over 2} (\nabla h)^2 +
{\alpha\over 2} m^2 + \zeta_h, \label{langevin1}\\ \partial_t m &=& K
(\nabla^2m+rm-um^3) + a\nabla h \cdot \nabla m+bm\nabla^2 h\nonumber \\
&& + {c\over 2} m (\nabla h)^2 + \zeta_m .\label{langevin2}
\end{eqnarray}
Here, we have included the lowest order (potentially relevant) terms
allowed by the symmetry $m \to -m$. 
Equation (\ref{langevin1})  is the Kardar-Parisi-Zhang (KPZ) equation\cite{kar86} 
for surface growth, plus a coupling to the order parameter.  
Equation (\ref{langevin2})  is the time dependent Landau--Ginzburg equation 
for a (non-conserved) Ising model, with three different couplings to the
height fluctuations. 
The Gaussian, delta-correlated noise terms, $\zeta_h$ and $\zeta_m$, 
mimic the effects of faster degrees of freedom. 
A different set of equations was proposed by 
L\'eonard and Desai\cite{leo97} for phase separation during MBE.  
Their equations reflect the MBE conditions of random particle
deposition (in contrast to sticking probabilities that depend on 
the local environment), and a conserved order parameter which
evolves by surface diffusion. 
They do not include the KPZ nonlinearity. 
Computer simulations of corresponding 1+1 dimensional systems 
are presented in \cite{leo97,leo97a}.

Dimensional analysis indicates that the couplings appearing in
Eqs.~(\ref{langevin1}-\ref{langevin2}) are relevant, and may lead to 
new universality classes.
We shall leave the renormalization group analysis of these equations
to a more technical paper, and focus here instead on computer
simulations in 1+1 dimensions. 
The quantities evaluated in the computer simulations are the height 
correlation function in Eq.~(\ref{Chh}), and the order parameter correlation 
functions perpendicular and parallel to the growth direction.
Allowing  for the possibility of different dynamic exponents,
$z_m$ and $z_h$, for the order parameter and the height variables,
we fit to the scaling forms
\begin{eqnarray}
G_m^{(x)}(x-x')&\equiv& \langle m(x,t) m( x',t) \rangle \nonumber\\
&=&|x-x'|^{\eta-1}g_m^{\perp}(| x-x'|/\xi)\nonumber\\ 
G_m^{(t)}(t-t') &\equiv& \langle m(x,t)m(x,t')\rangle\nonumber\\
&=& |t-t'|^{(\eta-1)/z_m} g_m^{\parallel}(|t-t'|/\xi^{z_m})\, . \label{corr}
\end{eqnarray}

\begin{figure}
\centerline{\epsfysize=0.06\columnwidth{{\epsfbox{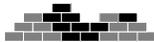}}}}
\narrowtext{\caption{The ``brick wall'' model used in the simulations.\label{fig1} 
}}
\end{figure}

Our simulations were done using a ``brick wall'' restricted
solid-on-solid model (see Fig.~\ref{fig1}).  
Starting from a flat surface, particles are added such that no overhangs 
are formed, and with the center of each particle atop the edge
of two particles in the layer below. 
We use two types of particles, $A$ and $B$ (black and grey in the figures). 
The probability for adding a particle to a given surface site, and the rule
for choosing its color, depend on the local neighborhood. 
When $A$ particles are more likely to be added to $A$ dominated regions,
and vice versa, the particles tend to phase separate and form domains. 
In this case, the order parameter correlation length $\xi$ is of the order of 
the average domain width. 
By changing the growth rules, it is possible to study cases in which 
some (or all) of the couplings $a$, $b$, $c$, and $\alpha$ vanish,
and thus to gain a complete picture of the different  ways in 
which the height and the order parameter influence each other. 

The decoupled case, $\alpha=a=b=c=0$, is implemented using the following
updating rules: A surface site is chosen at random, and a particle is 
added if it does not generate overhangs. 
Its color is then chosen depending on the colors of its two
neighbors in the layer below. 
If both neighbors have the same color, the newly added particle takes this
color with probability $1-p$, and the other color with probability $p$ 
(where $p$ is much smaller than 1). 
If the two neighbors have different colors, the new particle takes
either color with probability 1/2. 
Neighbors within the same layer are not considered.

\begin{figure}
\centerline{\epsfxsize=.97\columnwidth{{\epsfbox{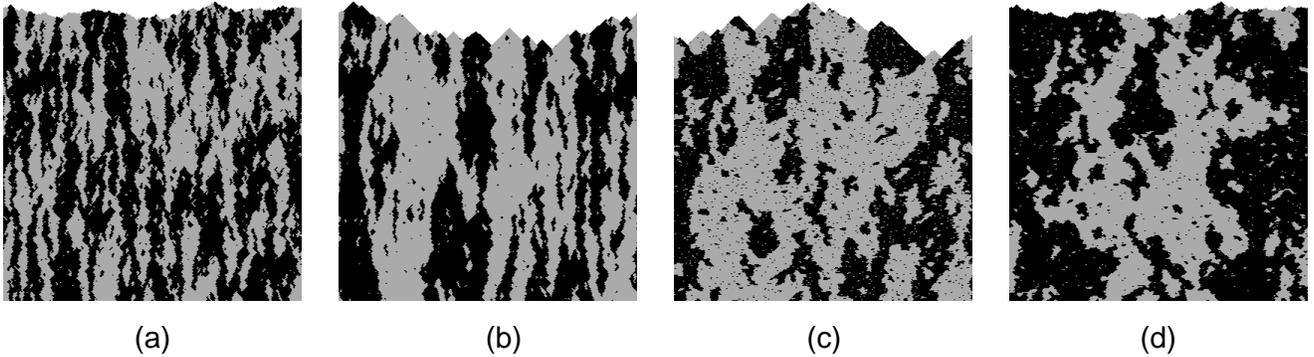}}}}
{\caption{Snapshot of the last 400 layers of simulations, for $L=200$ sites.
{\bf (a)} The decoupled case with $p=1/90$, and $r=1$.
{\bf (b)} For $p=1/200$, and $r=1/20$, the height is coupled to the domains, 
but not vice versa.
{\bf (c)} The fully coupled case, using the same parameters as (b), 
but with updating rules that include neighbors in the same layer.   
{\bf (d)} With $r=1$, and the updating rules of (c), the domains are
influenced by the height, but not vice versa. (Note that the profiles in
(a) and (d) are identical since we used the same random numbers.)
 \label{snaps}}}
\end{figure}

Since the probability of adding a particle to a given surface site
does not depend on its color, the surface grows exactly
 as with only one particle type, and is characterized by
the KPZ exponents $\chi=1/2$, and $z_h=3/2$. 
Similarly, the choice of particle color at a given
site is not affected by the height profile. 
The height profile determines only the moment at which a site is added, 
since the no-overhang condition requires both neighbors in the previous 
layer to be occupied. 
If we equate layer number with time, domain walls move to
the right or left with probability 1/2 during one time unit, and a
pair of new domain walls is created with probability $p$. 
This is identical to the Glauber model for  a 
one-dimensional Ising chain with coupling $J$ and at temperature $T$, 
with $p=\exp(4J/kT)$. 
The correlation length $\xi$ perpendicular to the
growth direction is consequently $\xi=\exp(-2J/kT)=1/\sqrt{p}$, 
and the correlation time is $\tau=\exp(-4J/kT)=1/p$. 
The dynamical critical exponent for the order parameter is thus $z_m=2$. 
Note that the ``time'' used for the order parameter (namely layer number)
is different from real time, which is for each particle the moment
when it is added to the growing surface. 
However, this difference becomes negligible for sufficiently small $p$ 
since the thickness of the surface over the correlation length, 
$\sqrt{\xi}$, is much smaller than the characteristic time, $\xi^2$, for
order parameter fluctuations.  
Simulations indeed confirm that the order parameter and height evolve 
completely independently.  
A typical profile is shown in  Fig.~\ref{snaps}a; the corresponding scaling
analysis conforms to expectations, and is not presented here.

The situation $\alpha>0$ with $a=b=c=0$ can be implemented by updating
sites on top of particles of different colors less often by a factor $r<1$ 
compared to sites above particles of the same color. 
As the order parameter is not affected by the height variable, 
its dynamics is still the same as that of an Ising model, with $z_m=2$. 
The height profile now has domain boundaries sitting preferentially
at its local minima, with mounds forming over domains (see Fig.~\ref{snaps}b). 
This leads to a surface roughness exponent of $\chi=1$ on length scales $\xi$, 
which is the case studied in\cite{kot98}. 
At these scales, changes in the height profile are slaved to domain wall motion,
and the dynamic exponent is $z_h=2$.
However, on length scales much larger than $\xi$, the KPZ exponents of
$\chi=1/2$ and $z_h=3/2$ are regained.
The crossover in the roughness can be described by a scaling form
$$ \langle[ h(x,t)-h(x',t)]^2\rangle = |x-x'|^2 g(|x-x'|/\xi),$$ 
with a constant $g(y)$ for $y\ll 1$, and $g(y) \sim 1/y$ for $y \gg 1$. 

To mimic the influence of surface roughness on the order parameter
(nonzero  $a$, $b$, or $c$ in Eqs.(\ref{langevin2})),
the color of a newly added particle is made dependent not only
on those of its two neighbors in the layer below, 
but also on the colors of its two nearest neighbors on the same layer, 
if these sites are already occupied. 
With probability $1-p$, the newly added particle takes the
color of the majority of its 2, 3, or 4 neighbors, and with
probability $p$ it assumes the opposite color. 
If there is a tie, the color is chosen at random with equal probability.
The height variable now affects the order parameter in two ways: 
{\bf (1)} {\it Domain walls are driven downhill.}
The reason is that the neighbor on the hillside of a site being updated 
is more likely to be occupied than the one on the valley side. 
The newly added particle is thus more likely to
have the color  on the hillside. 
(This corresponds to  $a>0$ in Eq.~(\ref{langevin2}).)
{\bf (2)} {\it New domains are predominantly formed on hilltops.} 
This is because domains on hilltops can expand more easily 
than those on slopes or in valleys, indicating  $b>0$ in Eq.~(\ref{langevin2}). 
Another consequence is that for the same  $p$,
the correlation length $\xi$ is much larger than in the 
decoupled case, as is apparent in Figs.\ref{snaps}c,d.

\end{multicols}\widetext
\begin{multicols}{2}\narrowtext

\begin{figure}
\centerline{\epsfxsize=0.7\columnwidth{{\epsfbox{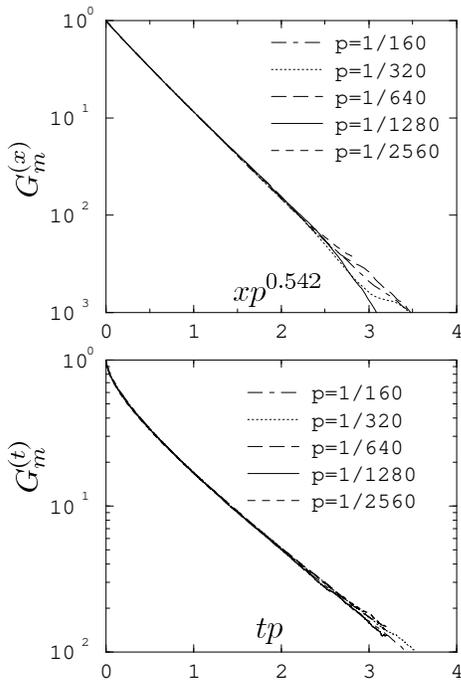}}}}
\narrowtext{\caption{Scaling collapse of correlations $G_m^{(x)}$ and
$G_m^{(t)}$ in Fig.~\protect{\ref{snaps}}d. 
For each $p$, the data is an average over
7500  widely separated layers, and for systems of size up to 8192.
\label{Gcollapse}}}
\end{figure}    

For the fully coupled case depicted in Fig.\ref{snaps}c we find essentially 
the same scaling behavior as in Fig.\ref{snaps}b, i.e. a height profile slaved
to the Glauber dynamics of the domains. 
The most interesting case, shown in Fig.\ref{snaps}d, is when the height
profile is independent of the domains ($\alpha=0$), evolving with KPZ dynamics,
while the order parameter is influenced by the roughness. 
The dynamic exponent $z_m$ for the order parameter was first 
obtained by collapsing the correlation functions using Eqs.~(\ref{corr}), 
as shown in Fig.\ref{Gcollapse}. 
These curves imply that $\eta = 1$, $\xi \propto p^{-0.542}$, 
and $\tau \propto \xi^{z_m} \propto 1/p$, 
giving $z_m \simeq 1/0.542 \simeq 1.85$. 

\begin{figure}
\centerline{\epsfysize=0.5\columnwidth{{\epsfbox{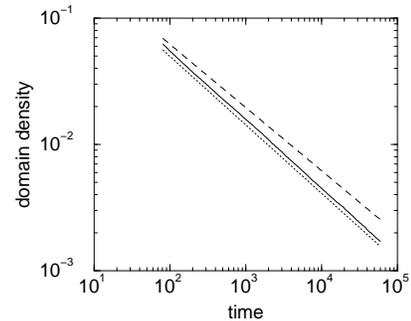}}}}
\narrowtext{\caption{Domain density as a function of time for $L=16384$,
averaged over 100 samples. The dotted line is a power-law fit
(slightly shifted for better visibility) with exponent of $1/z_m = 0.542$. 
For comparison, a power law with exponent $-0.5$ is also shown
(dashed line).
\label{fig7}}}
\end{figure}

The same non-trivial value for $z_m$ is obtained by a completely
independent measurement of the dynamics of domain coarsening 
following a quench from a ``high temperature'' ($p$ close to 0.5) 
to zero temperature ($p$=0).  
Fig.~\ref{fig7} shows the domain density as function of time 
for a system of size $L=16384$.
The resulting $z_m \simeq 1.85$, is in agreement
with the value from the scaling collapse.

The following simple argument fails to provide the exponent 
$z_m \simeq 1.85$.
Consider a Langevin equation,  $\dot x = \eta(t)$, for the position $x$
of a single domain wall at time $t$. 
Since the motion of the domain wall is strongly influenced by
the height profile, the noise $\eta(t)$ must have long-range correlations
$\langle \eta(t)\eta(t')\rangle = D |t-t'|^\alpha,$
reflecting the dynamics of surface.
This choice leads to $z_m=2$ for $\alpha >1$, 
and  $z_m=2/(2-\alpha)$ for $\alpha <1$. 
For a colored noise dominated by the slope fluctuations,
$\alpha=2/3$ and $z_m=3/2$, i.e. the height imposes its characteristic 
time scale  on  the order parameter. 
This would presumably be the case if the domain walls were
uniformly distributed along the surface.
However, due to their tendency to move downhill, 
they are preferentially found near valleys.
A different scaling of  the slope fluctuations in the valleys 
may be  the reason for the nontrivial value of $z_m$. 
Indeed, for short times, before the domain walls have moved 
to their preferred positions, the exponent $3/2$ is seen.

The dynamics of domain walls on a growing KPZ surface bears some
resemblance to the advection of a passive scalar in a turbulent
velocity field, which is characterized by nontrivial
exponents and multiscaling \cite{kra94}. 
If we neglect interactions between domain walls, and treat them as 
independent ``dust particles'' floating on the KPZ surface, 
the Langevin equation for the particle density $\rho$ is
\begin{equation}
\partial_t \rho = K \nabla^2\rho+ a(\nabla h \cdot \nabla \rho+\rho\nabla^2 h) + 
\zeta_\rho .\label{langevin3}
\end{equation}
The second term describes the {\it advection} of particles along a velocity
field $\vec v=\nabla h$.
Indeed this transformation maps the KPZ equation into the  Burgers equation 
for a vorticity-free, compressible fluid flow  \cite{kar86}. 
Equation (\ref{langevin3}) is a special case of Eq.~(\ref{langevin2}) for $m$,
with $r=u=c=0$, $b=a$, and with a conserved noise $\zeta_\rho$. 
(Together with Eq.~(\ref{langevin1}) for the height profile,  
it is also a special case of the equations used to 
describe the dynamic relaxation of drifting polymers\cite{ert93}.) 
In the remainder, we give the results of computer simulations for this case. 
The rules for the motion of ``dust particles'' are
identical to those for domain walls.
However, each particle is treated as if the others were not present. 
This means in particular that
there is no creation or annihilation of particles. 

\begin{figure}
\centerline{\epsfysize=0.55\columnwidth{{\epsfbox{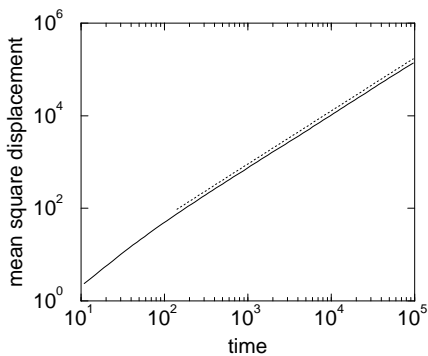}}}}
\narrowtext{\caption{Mean square displacement of a single domain wall in a 
system of size $L=4096$. The power law fit (dotted line) has an exponent 1.1467, 
corresponding to $z_\rho \simeq 1.74$. 
\label{fig8}}}
\end{figure}             

\begin{figure}
\centerline{\epsfysize=0.5\columnwidth{{\epsfbox{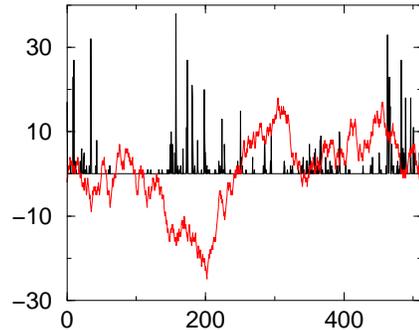}}}}
\narrowtext{\caption{
Histogram of the positions of 1024 domain walls
along a  surface profile (indicated in grey) of size $L=512$. 
\label{fig9}}}
\end{figure}

 Fig.~\ref{fig8} shows the mean square displacement of a single ``dust
particle'' in a system of size $L=4096$. 
To obtain good statistics, we averaged over 512 independent 
and noninteracting particles, and used more than 40 runs. 
The best fit is obtained for $z_\rho \simeq 1.74$, distinct from the 
previous $z_m \simeq 1.85$, implying that the exponents depend
on whether or not the  domain walls (or ``dust particles'') are conserved. 
In contrast to the advection of a passive scalar in a turbulent velocity field, 
we find no sign of multiscaling.
 Fig.~\ref{fig9} shows the positions of 1024 independent ``dust particles'' in 
a system of length $L=512$. 
While there is some correlation between minima of the surface profile 
and wall positions, there are also clusters of particles at higher elevations, 
indicating that particle diffusion is not sufficiently fast to fully adjust the
density to the faster changing height profile.
A fit of the density-density correlation  function to
$\langle \rho(x)\rho(0)\rangle\sim 1/x^{2(1-\chi_\rho)}$,
gives an exponent $\chi_\rho \simeq 0.85$.

In summary, the interplay between surface roughening and phase 
separation leads to a variety of novel critical scaling behaviors.
At one extreme, the height profile adapts to the dynamics of
critical domain ordering.
At the other, the dynamics of domain wall motion is influenced by
the roughness, exhibiting new and nontrivial scaling behaviors.

This work was supported by  EPSRC
(grant No.~GR/K79307, for BD), and
the National
Science Foundation (Grant No.  DMR-98-05833, for MK).

\end{multicols}
\end{document}